\documentclass[a4paper,12pt,showpacs,preprint,aip]{revtex4}

\usepackage{amsfonts}
\usepackage{rotating,psfig,epsfig}
\usepackage{amsmath,amssymb}
\usepackage{graphics,color}

\usepackage{fancyheadings}

\newtheorem{remark}{Remark}
  
\begin{document}

\title[Accumulation of embedded solitons]{Accumulation of embedded solitons \\ in systems with quadratic
nonlinearity}

\date{\today}

\author{B.~A.~Malomed}
\email{malomed@eng.tau.ac.il}
\affiliation{Department of Interdisciplinary Studies, Faculty of Engineering, Tel
Aviv University, Tel Aviv 69978, Israel}
\author{T.~Wagenknecht}
\email{t.wagenknecht@bristol.ac.uk}
\affiliation{Department of Engineering Mathematics, University of
Bristol, Bristol, BS8 1TR, UK}
\author{M.~J.~Pearce}
\email{m.j.pearce@lboro.ac.uk}
\affiliation{Department of Mathematical Sciences, Loughborough University, Loughborough,  LE11 3TU, UK}
\author{A.~R.~Champneys} 
\email{a.r.champneys@bristol.ac.uk}
\affiliation{Department of Engineering Mathematics, University of
Bristol, Bristol, BS8 1TR, UK}

\begin{abstract}
Previous numerical studies have revealed the existence of embedded
solitons (ESs) in a class of multi-wave systems with quadratic
nonlinearity, families of which seem to emerge from a critical point
in the parameter space, where the zero solution has a fourfold zero
eigenvalue. In this paper, the existence of such solutions is
studied in a three-wave model. An appropriate rescaling casts the
system in a normal form, which is universal for models
supporting ESs through quadratic nonlinearities. The normal-form
system contains a single irreducible parameter $\varepsilon $, and
is tantamount to the basic model of type-I
second-harmonic generation. An analytical approximation of WKB
type yields an asymptotic formula for the distribution of discrete
values of $\varepsilon $ at which the ESs exist. Comparison with
numerical results shows that the asymptotic formula yields an exact
value of the scaling index, $-6/5$, and a fairly good approximation
for the numerical factor in front of the scaling term.
\end{abstract}


\maketitle

\newpage

\pagestyle{fancy}
\setlength{\headrulewidth}{0pt}

\textbf{Soliton solutions in nonlinear models are parameterized by
their intrinsic frequency. An elementary principle determines
intervals of values of the frequency where ordinary solitons may
be found: these must lie in gaps (semi-infinite or finite ones) in which
the linearized version of the corresponding model has no solutions
for radiation waves. A usual argument supporting this principle is
that if one tries to construct a soliton with the frequency falling
within a band occupied by the radiation-wave solutions, the soliton
would start to decay by emitting linear waves whose frequency is
exactly equal to the soliton's intrinsic frequency. Nevertheless, it
has been demonstrated that exceptions are possible in practically
important models (notably, in a system of three waves coupled by
quadratic nonlinearity, with an extra linear coupling between two
fundamental-frequency waves induced by a Bragg grating, in an
underlying model of the spatial evolution of optical fields in a
planar waveguide). These exceptional solitons, which are
\textit{embedded} in the radiation spectrum, exist as a series of
isolated solutions at discrete values of the frequency inside the
radiation bands. An issue of considerable interest is the evolution
of these solutions following variation of governing parameters of
the model: in that sense, the embedded solitons form continuous
families. Previously, it was found in an empirical form (from direct
numerical results) that a presumably infinite series of 
embedded soliton solutions emanates from a single point in the
parameter plane of the above-mentioned three-wave system. 
The nature of this phenomenon was not understood.} 

\textbf{In this work, we resolve these issues, by developing an asymptotic 
analysis and verifying the predictions against numerical results. 
As a result, we derive a universal asymptotic approximation (\textit{normal form}) for
models with quadratic nonlinearity that support embedded solitons.
The normal form amounts to a system of two second-order equations,
which is well known as a basic model for ordinary solitons in 
second-harmonic-generating systems. There was a rather common belief 
that all soliton solutions had been found in this much-studied system. 
Nevertheless, in this work we are able to predict the existence of an infinite series of
previously unknown embedded solitons in the model. This is done in
an analytical form, by realizing that the embedded solitons feature
a broad inner zone, the solution in which must be matched to
exponentially decaying ones in outer zones. Further, a method
(Bohr-Sommerfeld quantization rule) borrowed from quantum mechanics
is applied to the solution in the inner zone. The most essential
part of the eventual analytical result is an asymptotic distribution
law for values of the single control parameter of the normal-form
system at which the embedded solitons exist. Comparison with
numerical results clearly shows that the analysis has produced an
\emph{exact} value of the respective scaling index.} %(it is
%}$-6/5$\textbf{).}

\section{Introduction}

This article deals with spatial solitary waves (or solitons in
short, without implying integrability of the underlying model) in a
model of a planar waveguide with a quadratic ($\chi ^{2}$)
nonlinearity, where two fundamental-frequency (FF) waves, $u_{1}$and
$u_{2}$, are linearly coupled by the Bragg reflection from a grating
of ridges or grooves on the surface of the waveguide. The waves are
coupled into the waveguide so that their amplitudes evolve along the
direction $z$, which is aligned with the Bragg grating, the Poynting
vectors of the two FF waves having angles $\pm \alpha $ with the $z$
axis (where the angle $\alpha $ is small). The $\chi ^{(2)}$-induced
nonlinear coupling between the FF waves generates a second-harmonic
(SH) wave, $u_{3}$. The corresponding dimensionless model, derived
in Ref. \cite{MaMaCh:98}, amounts to a system of normalized
equations,
\begin{equation} 
\begin{array}{rcl}
i(u_{1})_{z}+i(u_{1})_{x}+u_{2}+u_{3}u_{2}^{\ast } & = & 0, \\
i(u_{2})_{z}-i(u_{2})_{x}+u_{1}+u_{3}u_{1}^{\ast } & = & 0, \\
2i(u_{3})_{z}-qu_{3}+D(u_{3})_{xx}+u_{1}u_{2} & = & 0,\end{array}
\label{e:model}
\end{equation}
where $x$ is the transverse coordinate, the opposite signs in front of the
walk-off terms $\left( u_{1,2}\right) _{x}$ correspond to the
above-mentioned opposite angles $\pm \alpha $ ($\alpha $ itself does now
appear in the equations as it can be removed by a rescaling), the asterisk stands
for complex conjugation, and real coefficients $D$ and $q$ account for
the intrinsic diffraction and wave-number mismatch of the SH wave (the
intrinsic diffraction of the FF waves is neglected, as the artificial
diffraction, induced through the interplay of the Bragg-grating-induced
linear coupling and opposite walk-off terms, is stronger by several orders of
magnitude). Note that (\ref{e:model}) is invariant with respect to both a shift of phase 
by an arbitrary real constant $\phi $, 
\begin{equation}
S:(u_{1},u_{2},u_{3})\mapsto (\exp (i\phi )u_{1},\exp (i\phi )u_{2},\exp
(2i\phi )u_{3}),  \label{phi}
\end{equation}
and to an arbitrary transverse shift.

Solitary waves in Eqs. (\ref{e:model}) have been investigated in
several works \cite{MaMaCh:98,Mayer,ChMa:99b,Pe:03}. The present
work was motivated by results in the two latter references, which
had produced \emph{embedded solitons} (ESs) in the system. ESs are
distinguished by the feature that their internal frequency (spatial
frequency or propagation constant, in the present context) is in
resonance with radiation waves, \cite{ChMaYaKa:01}. In general, one
expects to find quasi-solitons with non-vanishing oscillating tails in such a
case. However, at certain discrete values of the frequency, the
amplitude of these tails can exactly vanish, thus giving rise to a
truly localized solution with a frequency that is \textit{embedded
in} the continuous spectrum.

One may seek for stationary ES solutions of (\ref{e:model}) in the form
\begin{equation}
u_{1,2}(x,z)=\exp (ikz)U_{1,2}(\xi ),\qquad u_{3}(x,z)=\exp (2ikz)U_{3}(\xi )
\label{xi}
\end{equation}
where $k$ denotes the wave number, and $\xi =x-cz$ is the tilted 
(``walking") coordinate, with $c$ measuring the walk-off of the spatial 
soliton's axis relative to the propagation direction $z$. Substitution of the
ansatz (\ref{xi}) into (\ref{e:model}) leads to a system of ordinary
differential equations for the complex amplitudes $U_{i}$,
\begin{equation}
\begin{array}{rcl}
-kU_{1}+i(1-c)U_{1}^{\prime }+U_{2}+U_{3}U_{2}^{\ast } & = & 0, \\
-kU_{2}-i(1+c)U_{2}^{\prime }+U_{1}+U_{3}U_{1}^{\ast } & = & 0, \\
-(4k+q)U_{3}+DU_{3}^{\prime \prime }-2icU_{3}^{\prime }+U_{1}U_{2} &
= & 0,\end{array} \label{e:8dim}
\end{equation}
with the prime standing for $d/d\xi $.

The analysis in this paper will concentrate on standing solitons
with $c=0$, in which case we can set $D=1$ with no loss of
generality. Furthermore, Eqs. (\ref{e:8dim}) with $c=0$ allow for 
symmetry reductions to four-dimensional invariant subspaces, 
the phase-shift-invariance (\ref{phi}) accounting for the existence of 
a one-parameter family of such invariant spaces. Localized solutions can be 
studied in any member of this family. A convenient reduction is achieved by 
setting $U_{1}=-U_{2}\equiv V\equiv V_{1}+iV_{2}$ ($V_{1}$ and 
$V_{2}$ are real), $U_{3}=U_{3}^{\ast }\equiv W$ (i.e., $W$ is real), which
reduces Eqs.~(\ref{e:8dim}) to a system
\begin{equation}
\begin{array}{rcl}
-kV+iV^{\prime }-VW-V^{\ast } & = & 0, \\
-(4k+q)W+W^{\prime \prime }-|V|^{2} & = & 0.\end{array}
\label{e:4dimc}
\end{equation}

A linear analysis reveals the region in parameter space where ESs
may be found. Indeed, in order for ESs to exist, the zero solution
of Eqs. (\ref{e:4dimc}), $V=W=0$, has to be a \emph{saddle-center},
i.e., the spectrum of the linearized system has to contain a pair of
real eigenvalues, such that solutions exponentially localized in one
direction exist, and a pair of purely imaginary eigenvalues, that
account, in the general case, for non vanishing tails in another
direction (if all the eigenvalues are real, the system gives rise to
ordinary gap solitons instead of ESs, \cite{ChMaYaKa:01}). For Eqs.
(\ref{e:4dimc}), it is straightforward to find that the zero
solution is a saddle-center in the region of the parameter space
determined by the inequalities
\begin{equation}
k^{2}<1,~4k+q<0,  \label{e:eslex}
\end{equation}see also Fig.~\ref{f:evdia}.

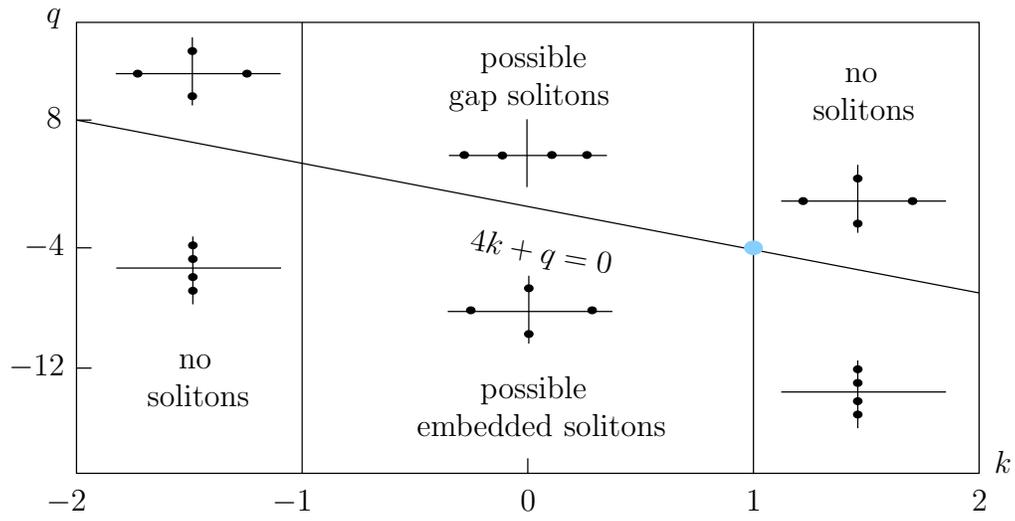
\begin{figure}
\caption{Enter Figure~\ref{f:evdia} here.}
\end{figure}

An important property of Eqs. (\ref{e:4dimc}) is \emph{reversibility}. This
means its invariance under the transformation (involution)
\begin{equation}
R_{1}:(V,W,W^{\prime })\mapsto (V^{\ast },W,-W^{\prime }),\quad \xi \mapsto
-\xi .  \label{reverse}
\end{equation}
An ES of Eqs. (\ref{e:model}) must be a solution of Eqs. (\ref{e:4dimc})
homoclinic to the saddle-center~$0$. For such a solution to exist,
the one-dimensional stable and unstable manifolds of the equilibrium
have to agree along the solution, and in general this situation is
of codimension three in $\mathbb{R}^{4}$. However, in an
$R_{1}$-reversible system obeying the invariance~(\ref{reverse}),
the stable and unstable manifolds of $0$ are $R_{1}$-images of one
another. Thus, a homoclinic orbit to $0$ exists if the unstable
manifold of $0$ intersects the two-dimensional fixed subspace of the
full four-dimensional phase space of Eqs. (\ref{e:4dimc}),
$\mathrm{Fix}(R_{1})=\{V_{2}=W^{\prime }=0\}$. This is a codimension-one 
situation and consequently, ESs should exist for parameter values on curves in the
two-dimensional parameter plane of the system (\ref{e:4dimc}).
Homoclinic orbits that intersect the fixed space of the involution
(\ref{reverse}) are themselves invariant under its action, therefore
they are called symmetric.

\begin{remark}
In addition, system (\ref{e:4dimc}) is reversible with respect to a second
involution,
\begin{equation}
R_{2}:(V,W,W^{\prime })\mapsto (-V^{\ast },W,-W^{\prime }),\quad \xi \mapsto
-\xi .
\end{equation}
According to the behavior of the first FF wave, $V_{1}$, we will call ESs
that are symmetric with respect to $R_{1}$ \textit{even} and those
symmetric with respect to $R_{2}$ odd. (Note that the SH ($W$)
component of both $R_{1} $- and $R_{2}$-symmetric homoclinic
solutions is automatically even.)
\end{remark}

\begin{remark}
This paper deals with fundamental ESs of (\ref{e:model}), i.e.,
one-pulse (single-humped) homoclinic solutions of Eqs.
(\ref{e:8dim}) or (\ref{e:4dimc}). It is well known that $N$-pulse
homoclinic orbits can emerge in bifurcations of homoclinic orbits to
saddle-center equilibria . In particular, it can be shown that
families of 2-pulse homoclinic orbits (bound-state solitons) must
accumulate on a family of ESs \cite{ChHa:00,MiHoOR:92}.
\end{remark}

%\subsection{Numerical results about embedded solitons}

It is straightforward to numerically construct both even and odd ESs, using
a variant of a numerical shooting method proposed in \cite{ChSp:93}, which 
searches for intersections of the unstable manifold
of $0$ with the corresponding fixed spaces $\mathrm{Fix}(R_{1,2})$.
Afterwards, branches of these solutions can been continued using the
software package AUTO \cite{Auto}. The corresponding results have already
been reported in Refs. \cite{ChMa:99b, Pe:03}.

Figure~\ref{f:bifdia} gives a concise overview of the computed
branches of ESs for Eqs. (\ref{e:4dimc}). In this paper, we are
interested in the situation close to the line $4k+q=0$. If this
curve is crossed (with $|k|<1$), the $0$ equilibrium changes its
type from a saddle-center into a saddle. It has been shown in Ref.
\cite{WaCh:03} that there is a possibility for curves of ES
solutions to approach this curve transversally. In that case, if the
curve is crossed, the soliton is not destroyed but rather changes
its type from embedded to an ordinary gap soliton. As can be seen in
Figure~\ref{f:bifdia}, this scenario \emph{does not} occur for
system~(\ref{e:4dimc}). Instead, one finds that all the curves of
the ES solutions approach the point $k=1,q=-4$, becoming
asymptotically tangent to the line $4k+q=0$. For these parameter
values $0$ is a degenerate codimension-two point, whose
linearization produces a fourfold zero eigenvalue. Our aim in this
paper is to understand how families of ESs accumulate at (emanate
from) this point.

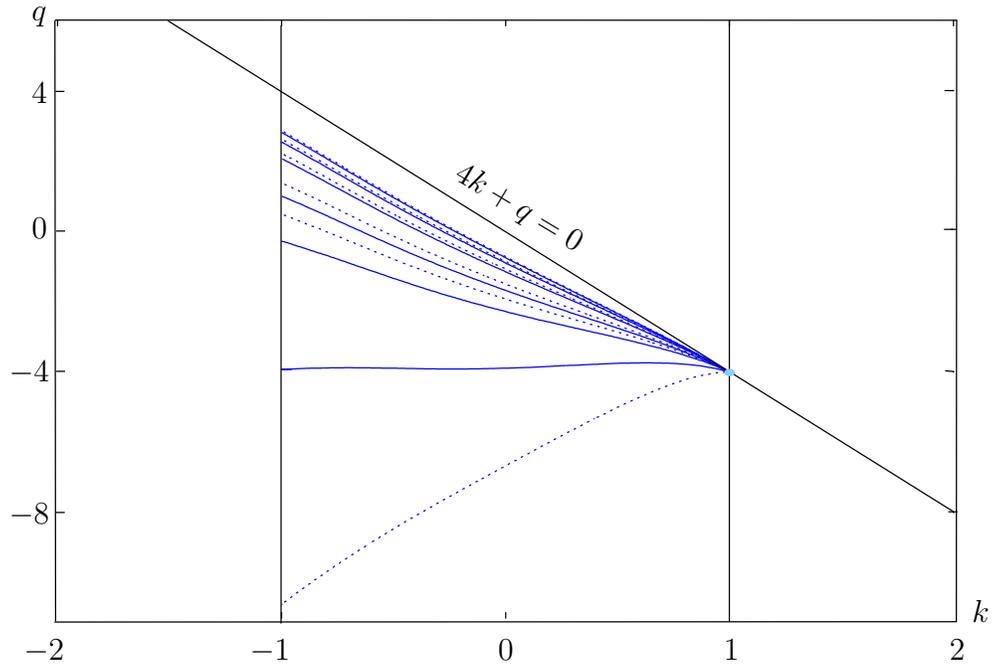
\begin{figure}
\caption{Enter Figure~\ref{f:bifdia} here.}
\end{figure}

In order to understand this accumulation we will perform an
asymptotic analysis of ES solutions. The analysis, however, will not
be performed directly for Eqs. (\ref{e:4dimc}). Instead, we perform
a rescaling of the original problem and derive a suitable normal
form in Section~\ref{s:rednf}. Afterwards, in Section~\ref{s:wkb},
an analytical WKB approximation yields asymptotic estimates for
parameter values at which ES solutions are predicted to exist. The
theory is verified by numerical results.

\section{The two-wave system as a normal form} \label{s:rednf}

The analysis of Eqs. (\ref{e:4dimc}) is facilitated by a rescaling of the
equations, suggested by the numerically computed solutions. First, we
display examples of such solutions. Fig.~\ref{f:solut} shows the FF and SH
components of ES solutions, which have been computed along the line $k=0$.
It can be seen that the FF component features internal oscillations, with
the number of oscillations increasing as the line $4k+q=0$ is approached. At
the same time, the SH component keeps a single-humped shape, with an
increasing amplitude.

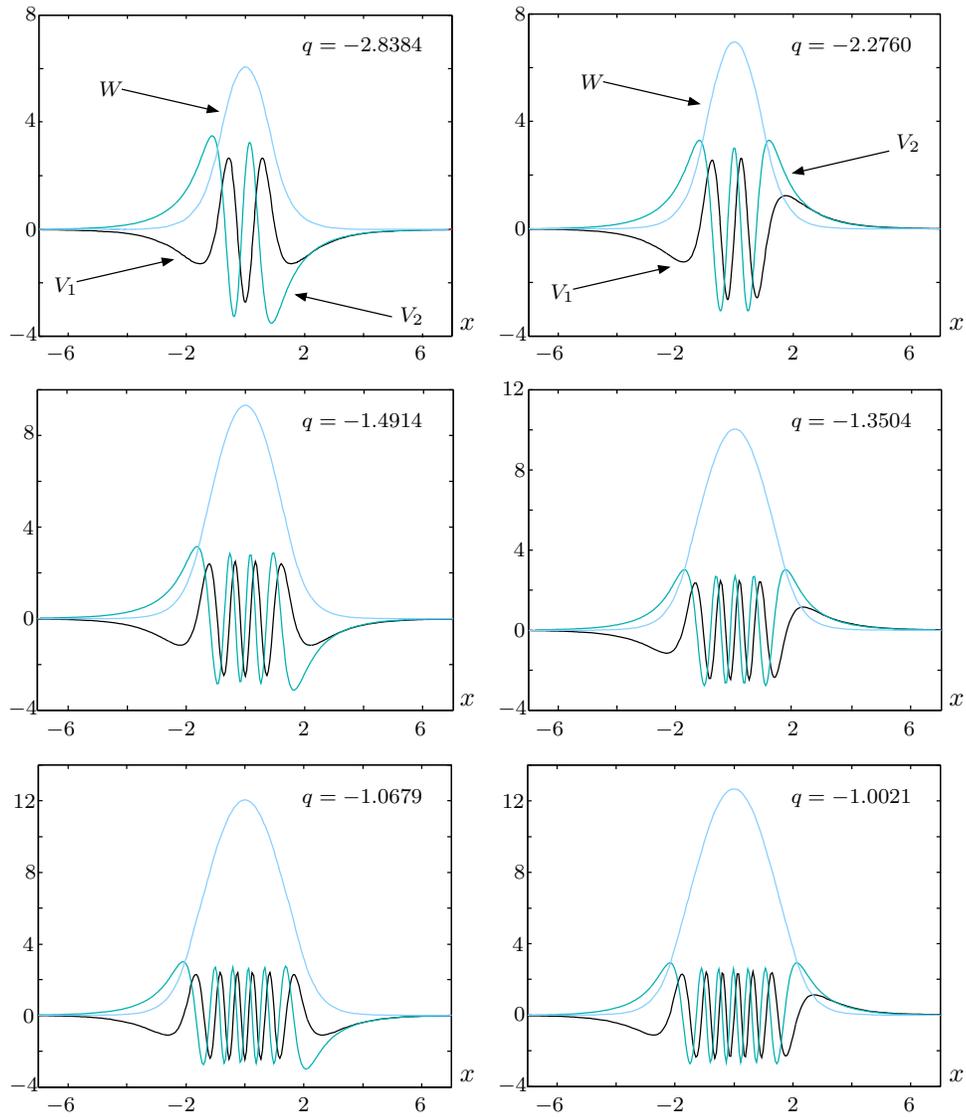
\begin{figure}
\caption{Enter Figure~\ref{f:solut} here.}
\end{figure}

It is interesting to interpret these numerical observations
geometrically. Note that geometric arguments have been successfully
used in Ref. \cite{WaCh:03} to understand the transition from ESs to
ordinary gap solitons (however, that work does not explain why this
transition is absent in the system (\ref{e:4dimc})).

Observe first that in the region $k^{2}<1$ and $4k+q<0$, where the
ESs exist, the $0$-equilibrium has a two-dimensional center manifold
$\mathcal{W}^{c}$, which is filled with periodic orbits. In fact,
from Eqs. (\ref{e:4dimc}) it can be seen that the flow in
$\mathcal{W}^{c}$ obeys the linear equation 
\[
W^{\prime \prime }=(4k+q)W, 
\]
which gives rise to periodic solutions with arbitrarily large
amplitude. This explains the (unusual) behavior of the ES solutions,
shown in Figure~\ref{f:solut}. More precisely, for different
branches of the ESs in Eqs. (\ref{e:4dimc}), the SH component $W$
grows indefinitely when the line $4k+q=0$ is approached. At the same 
time, the FF component, which corresponds to the hyperbolic direction 
of the $0$-equilibrium, shrinks and develops more and more internal
oscillations. Thus, in the $(V_{1},V_{2},W,W^{\prime })$ phase
space, the corresponding homoclinic orbit follows a periodic orbit
in $\mathcal{W}^{c}$, thereby oscillating around $\mathcal{W}^{c}$.
Both the size of the periodic orbit that is followed and the number
of oscillations around $\mathcal{W}^{c}$ increase as $4k+q=0$ is
approached.

Although these geometric considerations give insight into the behavior of
Eqs. (\ref{e:4dimc}), they do not explain the most remarkable feature, viz.,
the accumulation of branches of the ESs at $k=1$, $q=-4$. In fact, since the
$0$-equilibrium has a four-dimensional center manifold for these parameter
values, even the local behavior near this equilibrium is completely unknown
\textit{a priori}. We therefore address the existence of ES solutions via an
asymptotic analysis of a normal-form system derived below.

To this end, we again consider Eqs. (\ref{e:8dim}) with $c=0$, but
this time reduced to the invariant subspace $U_{1}=U_{2}^{\ast
}\equiv V$, $U_{3}=U_{3}^{\ast }=W$. Being interested in the
behavior near $k=1$ and $q=-4 $, we define $\alpha \equiv k-1$,
$\beta \equiv 4k+q$. Recall that due to the invariance of
(\ref{e:8dim}) with respect to $S$, the results will not depend on
the choice of a particular invariant subspace.

The equations for $V$ and $W$ then read
\begin{equation} \label{e:diffred}
\begin{array}{rcl}
-iV^{\prime }-\alpha V-(V-V^{\ast })+VW & = & 0, \\
W^{\prime \prime }-\beta W+|V|^{2} & = &
0.
\end{array}
\end{equation}
The scaling $\beta =\alpha \varepsilon $,
$\xi =(-\alpha )^{-1/2}x$, $V_{1}=(-\alpha )v_{1}$, $V_{2}=(-\alpha
)^{3/2}v_{2}$, and $W=(-\alpha )w$ transforms this system into
\begin{equation}
\begin{array}{rcl}
v_{1}^{\prime } & = & (2+\alpha )v_{2}+\alpha v_{2}w, \\
v_{2}^{\prime } & = & v_{1}+v_{1}w, \\
w^{\prime \prime } & = & -\varepsilon w-v_{1}^{2}+\alpha
v_{2}^{2},\end{array} \label{e:rescale}
\end{equation}where the prime now stands for $d/dx$. We are interested in solutions at $\alpha \ll 1$, for which $|v_{2}|\ll |v_{1}|$. In this case, Eqs.~(\ref{e:rescale}) can be further simplified to
\begin{equation}
\begin{array}{rcl}
v_{1}^{\prime } & = & 2v_{2}, \\
v_{2}^{\prime } & = & v_{1}+v_{1}w, \\
w^{\prime \prime } & = & -\varepsilon
w-v_{1}^{2},\end{array}\end{equation}or, setting $v\equiv v_{1}$, to
a system of two second-order equations,
\begin{equation}
\begin{array}{rcl}
v^{\prime \prime } & = & 2v+2vw, \\
w^{\prime \prime } & = & -\varepsilon w-v^{2}.
\end{array}
\label{e:2wave}
\end{equation}

System (\ref{e:2wave}) describes stationary solutions in optical
models with the $\chi ^{2}$ nonlinearity of the so-called type I,
which means that the system contains only two wave fields, rather than three. 
It has been studied in detail by a number of authors, see the reviews
\cite{Jena,BuTrSkTr:02}. The case we are interested in, with
$\varepsilon >0$, is generally classified as the ``bright-dark" case,
where the latter precisely means that it may have generic solutions
localized in one direction only. We view Eqs.~(\ref{e:2wave}) as a
\emph{normal form} for a class of models supporting embedded
solitons through quadratic nonlinearity.

\begin{remark}
Note that Eqs.~(\ref{e:2wave}) can also be interpreted in the context 
normal form theory, as, for example, introduced in \cite{Eletal:87}. Choosing
the coefficient in front of the linear term in the equation for $v$ to be another 
parameter, say $\mu$, instead of the constant value $2$, Eqs.~(\ref{e:2wave}) describe a normal form for
equilibria with fourfold eigenvalue zero in the class of reversible systems. It is 
interesting to note that the general normal form contains an additional
quadratic term $w^2$ in the equation for $w$. The fact that this term is missing 
in (\ref{e:2wave}) shows the special (degenerate) character
of the equations. (Note that this also implies a purely linear flow $w^{\prime \prime }  = 
-\varepsilon w$ in the centre manifold of the system.)  Consequently, former studies 
of normal forms near equilibria with fourfold eigenvalue zero in \cite{Io:92,Wa:02} do 
not discuss the accumulation of ESs observed in Eqs.~(\ref{e:2wave}).
\end{remark}

We shall now proceed with an asymptotic analysis to find a\emph{\
new class} of solutions to the much-studied equations
(\ref{e:2wave}); to the best of our knowledge, the numerous previous
studies of the system did not discuss the solutions that we report
on below. Note that the solutions have an unfeasibly large $w$
component and were probably missed because of that. Furthermore, as stationary 
solutions to the full type-I $\chi ^{(2)}$ model, these solitons are 
likely to be unstable, but they are, nevertheless, of importance for the 
understanding of the normal forms and intrinsic structure of the $\chi ^{(2)}$ 
models.

\section{Analysis of embedded solitons by means of the WKB approximation}
\label{s:wkb}

%The equations are
%\begin{eqnarray}
%\frac{d^{2}V}{dx^{2}} &=&2V+2VU_{3},  \label{V} \\
%\frac{d^{2}U_{3}}{dx^{2}} &=&-\varepsilon U_{3}-V^{2},  \label{U3}
%\end{eqnarray}
%where $\varepsilon$ is a positive parameter, and we are interested in the
%case $ \varepsilon \rightarrow +0$. In this case, following the pattern of
%the WKB (semi-classical) approximation in quantum mechanics, we
%look for a solution in the form:
%\begin{eqnarray}
%V &=&V_{0}(x)\sin \phi (x),~\frac{d\phi }{dx}\equiv k(x),  \label{phi} \\
%U_{3} &=&-U_{0}(x)+U_{2}(x)\cos \left( 2\phi (x)\right) ,  \label{U0}
%\end{eqnarray}
%where the functions $V_{0},U_{0},U_{2}$, and $k(x)$ are assumed to be slowly
%varying ones, while the phase $\phi (x)$ varies at a speed of order unity
%(i.e., it is a relatively rapidly varying function).

We are interested in the existence and shape of ES solutions of Eqs.
(\ref{e:2wave}) in the limit $\varepsilon \rightarrow +0$. After
applying the transformation to map solutions of Eqs. (\ref{e:diffred})
into those of Eqs. (\ref{e:2wave}), we find that in this limit the
$w$-component becomes large and wide, while the $v$-component
oscillates in an effective potential induced by $w(x)$ within this inner
core. Note that we focus on detecting even ES solutions of 
(\ref{e:2wave}), which are the fundamental ones.

\subsection{Inner zone analysis}

We first consider the core of the ES solution, that is, values of $x$, for which $w$ does not decay to $0$. Here we we shall look for solutions of Eqs.~(\ref{e:2wave}) via the ansatz 
\begin{equation}
\begin{array}{rcl}
v(x) & = & v_{0}(\delta x)\sin (\phi (x)), \\
w(x) & = & -w_{0}(\sqrt{\delta }x)+w_{2}(\delta x)\cos (2\phi
(x)).\end{array} \label{e:phi}
\end{equation}Here it is assumed that $\delta \ll 1$, such that $v_{0}$, $w_{0}$ and
$w_{2} $ are slowly varying functions.

Substitution of the ansatz (\ref{e:phi}) into Eqs. (\ref{e:2wave}) and
equating coefficients in front of $\sin \phi $ and $\cos \phi $ leads, in
the lowest-order approximation, to the following equations:
\begin{eqnarray}
{\phi ^{\prime }}^{2} &=&2\left( w_{0}-1\right) ,  \label{k} \\
\frac{d}{dx}\left( v_{0}^{2}\phi ^{\prime }\right) &=&0,\quad
\Longrightarrow v_{0}^{2}=\frac{2C}{\sqrt{w_{0}-1}},  \label{e:v0}
\end{eqnarray}
where $C$ is an arbitrary integration constant, and the factor $2$ was
introduced for convenience. Observe that Eq. (\ref{e:v0}) requires
$w_{0}>1$. Next, substituting Eqs. (\ref{e:phi}), and (\ref{k}),
(\ref{e:v0}) into the equation for $w$ in system (\ref{e:2wave}),
and equating coefficients as above, yields the following results:
\begin{equation}
\frac{d^{2}w_{0}}{dx^{2}}+\frac{\varepsilon }{\delta
}w_{0}-\frac{C}{\delta \sqrt{w_{0}-1}}=0,
\label{w0''}
\end{equation}
\begin{equation}
w_{2}=-\frac{v_{0}^{2}}{2\left( 4{\phi ^{\prime }}^{2}-\varepsilon
\right) }\equiv -\frac{C}{\sqrt{w_{0}-1}\cdot (8\left(
w_{0}-1\right) -\varepsilon )}.
\label{U2}
\end{equation}

%Note that the derivation of Eq. (\ref{k}) ignored a contribution from $U_{2}$
%. If that is taken into account, a corrected form of Eq. (\ref{k}) becomes
%\begin{equation}
%k^{2}=2\left( U_{0}-1\right) -\frac{\sqrt{2}C}{k\left( 4k^{2}-E\right) }.
%\label{corrected}
%\end{equation}
%Equation (\ref{corrected}) is, in principle, an algebraic equation
%for $k$ of the fifth degree. However, we will be actually
%interested in a case when $ C$ is small (see below), therefore a
%perturbative expansion of Eq. (\ref {corrected}) yields a small
%correction to the expression (\ref{k}):
%\begin{equation}
%k^{2}\approx 2\left( U_{0}-1\right) -\frac{C}{8\left( U_{0}-1\right) ^{3/2}}.
%\label{kcorrected}
%\end{equation}
%The approximation (\ref{kcorrected}) does not apply as $U_{0}-1\rightarrow 0$
%: in this case, Eq. (\ref{corrected}) yields, instead (when $U_{0}-1=0$
%exactly),
%\begin{equation}
%k^{3}\left( 4k^{2}-E\right) =-\sqrt{2}C.  \label{1/5}
%\end{equation}

After the rescaling $\xi \equiv \sqrt{\varepsilon /\delta }x$, Eq.
(\ref{w0''}) may be represented as
\begin{equation}
\frac{d^{2}w_{0}}{d\xi ^{2}}=-\frac{dP\left( w_{0}\right) }{dw_{0}},
\label{Newton}
\end{equation}i.e., as the equation of motion for a classical particle with mass
$m=1$ and the coordinate $w_{0}$, in a potential
\begin{equation}
P(w_{0})\equiv \frac{1}{2}\left( w_{0}^{2}-1\right) -K\sqrt{w_{0}-1},\quad
K\equiv \frac{2C}{\varepsilon }.  \label{potential}
\end{equation}

%\begin{equation}
%C\equiv \frac{1}{2}E\varepsilon ,  \label{varepsilon}
%\end{equation}
%the role of time being played by
%\begin{equation}
%\xi \equiv \sqrt{E}x  \label{xi}
%\end{equation}
%Generally, we are interested in the range of values $\varepsilon \sim 1$,
%hence Eq. (\ref{varepsilon}) shows that, with $E$ small, $C$ is small too,
%as it was already assumed above in Eq. (\ref{kcorrected}).

%A typical example of the effective potential (\ref{potential}) is shown in
%the figure [where the horizontal axis is $U_{0}-1$, and the vertical axis is
%$\left( 1/5\right) P\left( U_{0}\right) $, with $\varepsilon =5$].

%\FRAME{dtbpFX}{4.5in}{3in}{0pt}{}{}{Plot}{\special{language
%"Scientific Word";type "MAPLEPLOT";width 4.5in;height 3in;depth
%0pt;display "USEDEF";plot_snapshots TRUE;mustRecompute
%FALSE;lastEngine "MuPAD";xmin "-5";xmax "5";xviewmin
%"-0.004999999999999E0";xviewmax "5.005";yviewmin
%"-0.613694327705634";yviewmax "1.36590767319972";plottype
%4;numpoints 100;plotstyle "patch";axesstyle "normal";xis
%\TEXUX{x};yis \TEXUX{y};var1name \TEXUX{$x$};var2name
%\TEXUX{$y$};function
%\TEXUX{$0.1\left( x+1\right)^{2}-\sqrt{x}$};linecolor "black";linestyle 1;pointstyle
%"point";linethickness 1;lineAttributes "Solid";var1range
%"-5,5";num-x-gridlines 100;curveColor
%"[flat::RGB:0000000000]";curveStyle "Line";valid_file
%"T";tempfilename 'HTPOQS00.wmf';tempfile-properties "XPR";} }
In order for Eqs. (\ref{e:phi}) to describe ES solutions of Eqs.
(\ref{e:2wave}), which vanish at infinity, we must equate the
conserved energy of the mechanical model (\ref{Newton}) to zero,
$\left( 1/2\right) \left( {w_{0}^{\prime }}\right) ^{2}+P(w_{0})~=~0$.
This equation can be solved via the substitution
\begin{equation}
w_{0}(\xi )=1+\eta ^{4}(\xi ),  \label{eta}
\end{equation}which yields an implicit solution,
\begin{equation}
\xi =\int_{0}^{\eta _{0}}\frac{4\eta ^{2}}{\sqrt{2K-2\eta ^{2}-\eta
^{6}}}\; d\eta .  \label{e:eta}
\end{equation}Here, $\eta _{0}$ is a solution of
\begin{equation}
2K-2\eta ^{2}-\eta ^{6}=0.  \label{eta0}
\end{equation}
This implicit solution will be sufficient for the subsequent analysis.

%The potential always contains the potential well; if $\varepsilon $ is
%small, the well is small and shallow: in that case, its bottom
%point is located at $U_{0}-1\approx \varepsilon /2$, and the right
%edge (the point at which $P=0$) is at $U_{0}-1\approx \varepsilon$.

%After the substitution
%\begin{equation}
%U_{0}(\xi )\equiv 1+\eta ^{4}(\xi ),  \label{eta}
%\end{equation}
%a solution to Eq. (\ref{Newton}) can be represented in an implicit form, as
%it follows from the conservation of the mechanical energy, $(1/2)\left(
%dU_{0}/d\xi \right) ^{2}+P\left( U_{0}\right) =0$:
%\begin{equation}
%\xi =\int_{0}^{\eta }\frac{\left( \eta ^{\prime }\right)
%^{2}}{\sqrt{ 2\varepsilon -2\left( \eta ^{\prime }\right)
%^{2}-\left( \eta ^{\prime }\right) ^{6}}}d\eta ^{\prime },
%\label{implicit}
%\end{equation}
%where the integration extends up to the point $\eta _{0}$, which is a root
%of the equation
%\begin{equation}
%2\varepsilon -2\eta _{0}^{2}-\eta _{0}^{6}=0  \label{eta0}
%\end{equation}
%($\eta _{0}\approx \sqrt{\varepsilon }$, if $\varepsilon \ll 1$).

\subsection{The Bohr-Sommerfeld quantization of the constant $C$}

Using the above results, we can apply the \textit{Bohr-Sommerfeld
}(BS)\textit{\ }quantization rule (a part of the semi-classical or
WKB approximation in quantum mechanics), which selects a
discrete spectrum of values of $\varepsilon $, necessary for the
existence of the ES solution. Recall that the analysis so far has
been valid for the case of $w_{0}>1$. If $w_{0}=1$, then $\phi
^{\prime }=0$, because of Eq. (\ref{k}), and we face a turning-point
problem. Moreover, $w_{0}(x)=1$ leads to a divergence in Eq.
(\ref{e:v0}), which has to be compensated by setting $\sin (\phi
(x))=0$. Obviously, there are two values of $x$ for which
$w_{0}(x)-1$ vanishes. In fact, these points describe the transition
from the core of the ES to its asymptotic tails, where the solution
decays to $0$. At one of the two points, say $x=0$, we simply set
$\phi =0$. Then the condition that $\sin \phi $ also vanishes at the
second point, $x=X$, implies that
\begin{equation}
\phi (X)=\int_{0}^{X}\phi ^{\prime }(x)dx=n \pi,~n=1,2,3,...
\end{equation}Observe that $n$ counts the number of ``inner oscillations"
of the $v$-component of the derived solution. We thus obtain
\begin{eqnarray}
n\pi &=&\int_{0}^{\sqrt{\varepsilon }X}\sqrt{\frac{2}{\varepsilon }} \;\; \eta
^{2}(\xi )\;d\xi  \notag \\
&=&\int_{0}^{\; \eta _{0}}\sqrt{\frac{2}{\varepsilon }}\; \frac{4\eta
^{4}}{\sqrt{2K-2\eta ^{2}-\eta ^{6}}}\; d\eta .  \label{Bohr}
\end{eqnarray}
Equation (\ref{Bohr}) is tantamount to the BS rule in the semi-classical limit
of quantum mechanics.
%$(in fact, it is valid only for sufficiently large values of the integer $n$
%$).

We now have to distinguish two cases. First we assume $K\ll 1$. In this
case we obtain from Eq. (\ref{eta0}) that $\eta _{0}\approx \sqrt{K}$, and
the BS rule yields
\begin{equation}
\sqrt{\frac{2}{\varepsilon }}\int_{0}^{1}\sqrt{K}\frac{4K^{2}\tau
^{4}}{\sqrt{2K-2K\tau ^{2}-K^{3}\tau ^{6}}}\;\; d \tau =n\pi .  \label{BS}
\end{equation}In particular, for small $\varepsilon $, making use of the fact that
\begin{equation}
\int_{0}^{1}\frac{\eta ^{4}d\eta }{\sqrt{1-\eta ^{2}}}=\allowbreak
\frac{3}{16}\pi ,
\end{equation}the BS quantization rule takes an eventual explicit form,
\begin{equation}
K_{n}^{2}=\frac{4}{3}\sqrt{\varepsilon }n.  \label{final}
\end{equation}
On the other hand, for large $K\gg 1$, we have $\eta _{0}={2K}^{1/6}$. In a
similar way, using the numerical value
\begin{equation}
\int_{0}^{1}\frac{\tau ^{4}}{\sqrt{1-\tau ^{6}}}\;d\tau =\frac{\Gamma \left(
2/3\right) \Gamma \left( 5/6\right) }{4\sqrt{\pi /3}}\approx \allowbreak
0.373,
\end{equation}the BS rule leads to
\begin{equation}
\left( 2K_{n}\right) ^{1/3}=\frac{n\pi }{4\sqrt{2}\cdot
0.373}\sqrt{\varepsilon }.  \label{large-varepsilon}
\end{equation}

\subsection{The asymptotic solution in the outer zone, and quantization of $\protect\varepsilon$}

The above analysis has been dealing with the \textit{inner zone} (core)
of the ES solution, where $w_{0}$ took values $w_{0}\geq 1$. It remains
to match the results to the solution in the \textit{outer zone} (tail),
where $w_{0}$ decays to $0$. To this end, we note first that the solution
for the fundamental field, which takes the form of Eq. (\ref{e:phi}) in the
inner zone, can be matched to an exponentially decaying solution for the
same field in the outer zone, given by
\begin{equation}
v_{\mathrm{outer}}(x)=A\exp \left( -\sqrt{2}|x|\right) .  \label{outer}
\end{equation}The actual issue is to relate the constant $A$ in this expression to $C$ in
Eq. (\ref{e:v0}). This can be done using the known \emph{connection formula}
\cite{BeOr:78} of the WKB approximation, with the result
\begin{equation}
A=\sqrt{C/2}.  \label{A}
\end{equation}

In the same outer zone, we can use the fact that the ES is
a localized solution. This implies that the corresponding solution for the
SH field is
\begin{equation}
w(x)\approx B\exp \left( -2\sqrt{2}|x|\right) .  \label{U3-outer}
\end{equation}The amplitude $B$ can be computed using the differential equation
(\ref{e:2wave}) for $w$ and relation (\ref{A}). We thus find
\begin{equation}
B=-A^{2}/8=-C/16.  \label{B}
\end{equation}

Now, we can match the outer and inner solutions for the SH field,
demanding continuity of this field. According to the above results,
the inner-field solution takes the value $\left( w\right)
_{\mathrm{inner}}(x=0)=-w_{0}(x=0)=-1$, see again (\ref{e:phi}), at
the border between the two zones. The matching condition thus
implies $C=16$, and therefore we find
\begin{equation}
\varepsilon K_{n}=32.  \label{varepsilonE}
\end{equation}

It should be noted that this matching procedure is not a rigorous one, which
would demand to construct an intermediate asymptotic solution in a transient
zone, whose functional form (rather than just values of the fields at the
contact points) must be matched to ones in the inner and outer zones. Such a
rigorous procedure would be extremely complicated in the present system,
while the simpler one described above yields, eventually, rather accurate
results, as shown below.

We can now use the ``quantization rules" (\ref{final}) and
(\ref{large-varepsilon}) to derive approximations for values
$\varepsilon _{n} $ at which ES solutions exist in
system~(\ref{e:2wave}). We again start with the case $K\ll 1$. The
above considerations show that
\begin{equation}
\varepsilon
_{n}=\frac{32}{K_{n}}=\frac{16\sqrt{3}}{{\sqrt{\varepsilon
_{n}}n}^{1/2}},
\end{equation}and therefore,
\begin{equation}
K_{n}=\frac{32}{(16\sqrt{3})^{4/5}}\cdot n^{2/5}.
\end{equation}However, this implies $K_{n}\rightarrow \infty $ for $n\rightarrow \infty $,
in contrast to $K\ll 1$. So this case is irrelevant.

In the opposite case, when $K$ is large, the combination of Eq.
(\ref{varepsilonE}) and the corresponding approximation
(\ref{large-varepsilon}) produces the result\begin{equation}
\varepsilon _{n}=3.27\cdot n^{-6/5}.  \label{finform}
\end{equation}The applicability of this result also demands $n$ to be large enough. The
condition $K\gg 1$, which was used to derive Eq. (\ref{finform}),
now reduces to $32/\varepsilon _{n}\approx 9.8\cdot n^{6/5}\gg 1$,
which is \emph{always satisfied}. So, Eq. (\ref{finform}) yields an
asymptotic formula for values of $\varepsilon _{n}$ at which Eqs.
(\ref{e:2wave}) support ES solutions.

%\label{s:numeric}

\subsection{Numerical verification}

We now investigate the validity of the analytical prediction result
(\ref{finform}) through numerical solution of Eqs. (\ref{e:2wave}).
Similar to the case of system (\ref{e:4dimc}), ES solutions can be
computed by searching intersections of the unstable manifold of $0$
with $\mathrm{Fix}(R)$ by dint of the shooting method. This way, the
first 140 solutions have been constructed. The results are
summarized in Fig.~\ref{f:matt_fig}. In the main panel e) we plot
$n^{-6/5}$ against $\varepsilon $, where $n$ is identified as the
number of oscillations of the $v$-component in the solution's
inner zone, following the consideration of the soliton's structure
presented above in Section~\ref{s:wkb}. For $n=1,2,10,20$, the
computed solutions are shown in panels a) - d). Panel~f) specially displays a zoom of
the small square from e), which corresponds to $50 \leq n \leq 140$. For these values
of $n$ we may reasonably assume 
may Eq. (\ref{finform}) to be applicable. Again, the axes are chosen
to be $n^{-6/5}$ and $\varepsilon $ with the objective to compare the numerical 
and analytical results.

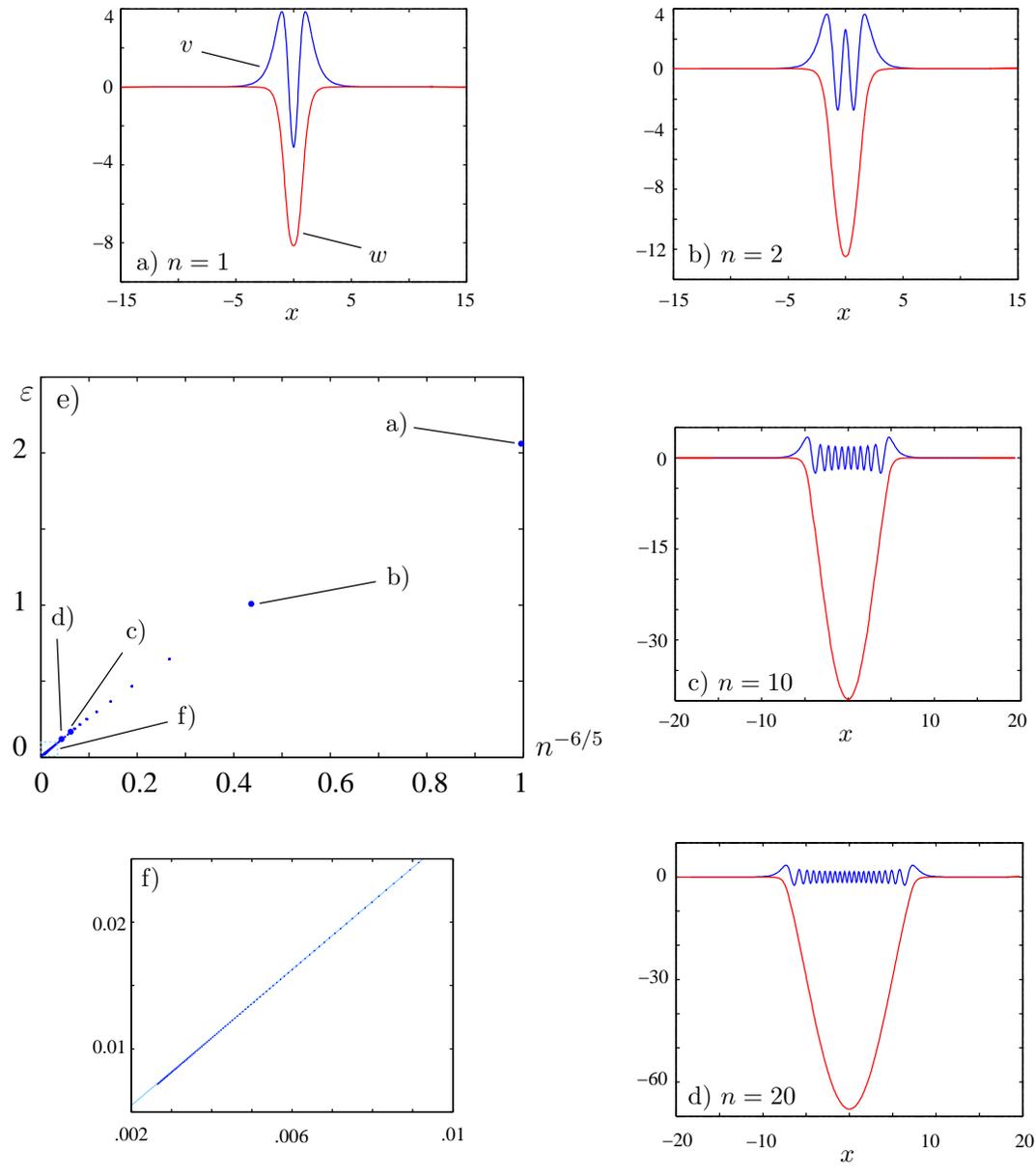
\begin{figure}
\caption{Enter Figure~\ref{f:matt_fig} here.}
\end{figure}

A linear regression analysis for the close-up in panel f) yields the
following fit to the computed values,
\begin{equation}
\varepsilon _{n}=6\cdot 10^{-5}+2.8247\cdot n^{-6/5},  \label{fit}
\end{equation}which is also shown in Fig. \ref{f:matt_fig}(f). First of all, it
corroborates the main result of the above asymptotic analysis, viz.,
that the \textit{scaling index} for $\varepsilon _{n}$ is $-6/5$.
Further, the numerical factor in front of $n^{-6/5}$, predicted by
Eq. (\ref{f:matt_fig}), and its empirical counterpart in Eq.
(\ref{fit}) differ by $\simeq 15\%$. This moderate discrepancy may
be explained by the insufficiently accurate matching procedure, as
discussed above. Finally, the very small constant term in Eq.
(\ref{fit}) is actually caused by the contribution from the ES
solutions with smaller $n$, to which the asymptotic analysis does
not apply. In particular, limiting the numerical data to $80 \leq n \leq 140$
reduces the constant term by a factor of $6$, and shows that all the
data fall into a $95\%$--confidence interval corresponding to a
strictly proportional relation
between $\varepsilon _{n}$ and $n^{-6/5}$, without the constant term in Eq. (\ref{fit}).

\section{Conclusions}

In this paper, we have derived a system of two second-order equations
with one free parameter, $\varepsilon $, as an asymptotic normal form
of models with quadratic nonlinearity, close to a critical point from
which branches of embedded-soliton (ES)\ solutions originate. At the
critical point the zero solution possesses a fourfold eigenvalue zero.

The derived normal form equations are tantamount to a well-known
system, which is fundamental for the usual second-harmonic-generation
model of type I.  Despite the fact that the system has been much
studied, we were able to find a new infinite series of ES solutions in
this work. A fundamental characteristic of the series is an asymptotic
distribution law for discrete values of $\varepsilon $ at which the ES
exist. Making use of the Bohr-Sommerfeld quantization rule borrowed
from the semi-classical version of quantum mechanics and a simplified
procedure of matching the solutions in the inner and outer zones of
the soliton we were able to predict the asymptotic distribution law in
a fully analytical form. The comparison with direct numerical results
shows that the predicted scaling law, $\varepsilon
_{n}=\mathrm{const}\cdot n^{-6/5}$, may be absolutely exact (no
discrepancy in the value of the scaling index, $-6/5$, was found), and
the constant factor was predicted with an error $\simeq 15\%$ (due to
an inaccuracy of the simplified matching procedure).

These results present fundamental (and previously unknown in any form)
information about families of ES solutions in multi-wave systems with
quadratic nonlinearity. Additionally, they shed new light on the
above-mentioned system~(\ref{e:2wave}) that describes type-I
second-harmonic generation.

It is worthwhile to contrast the features of the families of ESs that
we have established with other infinite families of ESs found
in the literature, e.g.~\cite{ChMaYaKa:01,YaAk:03,KoChBuSa:02}.
In all other cases we are aware of
the individual members of the family correspond to
multi-humped solitary waves. In contrast, our construction produces
waves that are fundamental in one component and become highly
oscillatory in the second component. Effectively these are single-humped
waves. 

There are chiefly two mechanism
that produce multi-humped waves, under a so-called Birkhoff
signature condition on the sign of the nonlinear term. This condition
is satisfied by the three-wave system investigated here. One theory
(see remark 2 above) shows that there must be an infinite family of
branches of multi-humped ESs accumulating on each branch of
single-humped ES.  The other theory \cite{KoChBuSa:02} shows that
multi-humped waves must accumulate on the limit corresponding to $k=1$
in this model. Further numerical results not presented here
\cite{Pe:03} compute representatives of both kinds of multi-humped ESs
and follow their branches in the $(k,q)$-plane.  Note that the latter kind
also produces a family of curves that originate from the point
$q=-4,k=1$ in Figure~2 \cite[Fig 4.3]{Pe:03}
similar to the families computed by Yang and
Akylas \cite{YaAk:03} for the third-order nonlinear Schr\"{o}dinger
equation. The aim of this paper has been to establish why in addition,
an infinite family of {\em single-humped} solitons should emanate from
the same critical point in the three-wave model.

We have already pointed out that we do not expect these newly-established
families of excited ESs to be stable for the full PDE system.
This has been demonstrated in
numerical studies for the 3-wave system~(\ref{e:model}) in
\cite{Pe:03}. Nevertheless, it has also been shown there that if the
parameters are chosen close enough to the critical value, the even
ground-state solution shown in the top panel of Figure~\ref{f:solut}
is {\em virtually stable}. This means that although the central cores
emit radiation tails, those have an extremely small amplitude and it
takes a considerable distance for the soliton solution to show
instability. We would expect that the excited states would also develop
very slow instabilities near to the critical point. Thus, there may be
physical application of these multitude of states, for example in optical
memory, as in most experimental systems the residence time is much
smaller than that required to produce the instability.

\section*{Acknowledgement}

B.A.M. appreciates hospitality of the Department of Engineering Mathematics
at the University of Bristol under support of EPSRC grant GR/R72020/01 (BCANM). 
T.W. acknowledges support by EPSRC grant GR/535684/01.

%\bibliographystyle{unsrt}
%\bibliography{thesis}

\newpage

\section*{List of Captions}

\setcounter{figure}{0}

\begin{figure}[h]
\caption{Results of the linear analysis predicting the possible
existence of embedded solitons in the case of $D=1$. In each part of
the parameter plane, the set of eigenvalues found from the equations
(\protect\ref{e:4dimc}) linearized around the zero solution is
displayed.The large dot at $(k,q)=(1,-4)$ denotes a codimension-two
point at which the linearized system has a fourfold eigenvalue zero.
\label{f:evdia}}
\end{figure}

\begin{figure}[h]
\caption{Branches of even (dotted curves) and odd (solid curves) ESs in
system (\protect\ref{e:4dimc}), as per Ref. \protect\cite{Pe:03}. No branch
crosses the line $4k+q=0$, but all of them approach the point $(k,q)=(1,-4)$
tangentially to this line. \label{f:bifdia}}
\end{figure}

\begin{figure}[h]
\caption{Examples of even (left panels) and odd (right panels)
embedded-soliton solutions of Eqs. (\protect\ref{e:4dimc}), computed
for $k=0 $, as per Ref. \protect\cite{Pe:03}. Recall that $V_{1}$
and $V_{2}$ are the real and imaginary parts of the complex field
$V$. \label{f:solut}}
\end{figure}

\begin{figure}[!h]
\caption{Panels a) through d) show characteristic examples of the
embedded solitons found from the numerical solution of Eqs.
(\protect\ref{e:2wave}). $140$ values of $\protect\varepsilon $ at
which the embedded solitons are found are collected in the main box
e), $n$ being identified as the number of oscillations of the
$v$-component in the soliton's inner zone. Panel f) is a close-up of
a part of the main diagram for $50<n<140$. It also shows the
straight line (\protect\ref{fit}) which provides for the best fit to
the data included in this panel. \label{f:matt_fig}}
\end{figure}

\newpage

\section*{Figure 1}

\setcounter{figure}{0}

\vspace*{3cm}

\begin{figure}[h]
\begin{center}
\begin{picture}(0,0)%
\includegraphics{malomed_fig1.pstex}%
\end{picture}%
\setlength{\unitlength}{4144sp}%
\begingroup\makeatletter\ifx\SetFigFont\undefined%
\gdef\SetFigFont#1#2#3#4#5{%
  \reset@font\fontsize{#1}{#2pt}%
  \fontfamily{#3}\fontseries{#4}\fontshape{#5}%
  \selectfont}%
\fi\endgroup%
\begin{picture}(5895,3130)(-404,-5735)
\put(5491,-5461){\makebox(0,0)[lb]{\smash{\SetFigFont{12}{14.4}{\familydefault}{\mddefault}{\updefault}{\color[rgb]{0,0,0}$k$}%
}}}
\put(2038,-5236){\makebox(0,0)[lb]{\smash{\SetFigFont{12}{14.4}{\familydefault}{\mddefault}{\updefault}{\color[rgb]{0,0,0}embedded solitons}%
}}}
\put(2417,-5011){\makebox(0,0)[lb]{\smash{\SetFigFont{12}{14.4}{\familydefault}{\mddefault}{\updefault}{\color[rgb]{0,0,0}possible}%
}}}
\put(2227,-3256){\makebox(0,0)[lb]{\smash{\SetFigFont{12}{14.4}{\familydefault}{\mddefault}{\updefault}{\color[rgb]{0,0,0}gap solitons}%
}}}
\put(2417,-3031){\makebox(0,0)[lb]{\smash{\SetFigFont{12}{14.4}{\familydefault}{\mddefault}{\updefault}{\color[rgb]{0,0,0}possible}%
}}}
\put(4596,-3121){\makebox(0,0)[lb]{\smash{\SetFigFont{12}{14.4}{\familydefault}{\mddefault}{\updefault}{\color[rgb]{0,0,0}no}%
}}}
\put(4406,-3346){\makebox(0,0)[lb]{\smash{\SetFigFont{12}{14.4}{\familydefault}{\mddefault}{\updefault}{\color[rgb]{0,0,0}solitons}%
}}}
\put(612,-4831){\makebox(0,0)[lb]{\smash{\SetFigFont{12}{14.4}{\familydefault}{\mddefault}{\updefault}{\color[rgb]{0,0,0}no}%
}}}
\put(423,-5056){\makebox(0,0)[lb]{\smash{\SetFigFont{12}{14.4}{\familydefault}{\mddefault}{\updefault}{\color[rgb]{0,0,0}solitons}%
}}}
\put(2349,-4119){\rotatebox{350.0}{\makebox(0,0)[lb]{\smash{\SetFigFont{12}{14.4}{\familydefault}{\mddefault}{\updefault}{\color[rgb]{0,0,0}$4k+q=0$}%
}}}}
\put(-179,-2761){\makebox(0,0)[lb]{\smash{\SetFigFont{12}{14.4}{\familydefault}{\mddefault}{\updefault}{\color[rgb]{0,0,0}$q$}%
}}}
\put(5356,-5686){\makebox(0,0)[lb]{\smash{\SetFigFont{12}{14.4}{\familydefault}{\mddefault}{\updefault}{\color[rgb]{0,0,0}$2$}%
}}}
\put(1171,-5686){\makebox(0,0)[lb]{\smash{\SetFigFont{12}{14.4}{\familydefault}{\mddefault}{\updefault}{\color[rgb]{0,0,0}$-1$}%
}}}
\put(-179,-5686){\makebox(0,0)[lb]{\smash{\SetFigFont{12}{14.4}{\familydefault}{\mddefault}{\updefault}{\color[rgb]{0,0,0}$-2$}%
}}}
\put(-179,-3391){\makebox(0,0)[lb]{\smash{\SetFigFont{12}{14.4}{\familydefault}{\mddefault}{\updefault}{\color[rgb]{0,0,0}$8$}%
}}}
\put(-314,-4156){\makebox(0,0)[lb]{\smash{\SetFigFont{12}{14.4}{\familydefault}{\mddefault}{\updefault}{\color[rgb]{0,0,0}$-4$}%
}}}
\put(-404,-4876){\makebox(0,0)[lb]{\smash{\SetFigFont{12}{14.4}{\familydefault}{\mddefault}{\updefault}{\color[rgb]{0,0,0}$-12$}%
}}}
\put(2656,-5686){\makebox(0,0)[lb]{\smash{\SetFigFont{12}{14.4}{\familydefault}{\mddefault}{\updefault}{\color[rgb]{0,0,0}$0$}%
}}}
\put(4006,-5686){\makebox(0,0)[lb]{\smash{\SetFigFont{12}{14.4}{\familydefault}{\mddefault}{\updefault}{\color[rgb]{0,0,0}$1$}%
}}}
\end{picture}

\end{center}
\end{figure}

\newpage

\section*{Figure 2}

\vspace*{3cm}

\begin{figure}[h]
\begin{center}
\begin{picture}(0,0)%
\includegraphics{malomed_fig2.pstex}%
\end{picture}%
\setlength{\unitlength}{4144sp}%
\begingroup\makeatletter\ifx\SetFigFont\undefined%
\gdef\SetFigFont#1#2#3#4#5{%
  \reset@font\fontsize{#1}{#2pt}%
  \fontfamily{#3}\fontseries{#4}\fontshape{#5}%
  \selectfont}%
\fi\endgroup%
\begin{picture}(5760,4030)(-269,-4835)
\put(2386,-1906){\rotatebox{329.0}{\makebox(0,0)[lb]{\smash{\SetFigFont{12}{14.4}{\familydefault}{\mddefault}{\updefault}{\color[rgb]{0,0,0}$4k+q=0$}%
}}}}
\put(5491,-4561){\makebox(0,0)[lb]{\smash{\SetFigFont{12}{14.4}{\familydefault}{\mddefault}{\updefault}{\color[rgb]{0,0,0}$k$}%
}}}
\put(2656,-4786){\makebox(0,0)[lb]{\smash{\SetFigFont{12}{14.4}{\familydefault}{\mddefault}{\updefault}{\color[rgb]{0,0,0}$0$}%
}}}
\put(1171,-4786){\makebox(0,0)[lb]{\smash{\SetFigFont{12}{14.4}{\familydefault}{\mddefault}{\updefault}{\color[rgb]{0,0,0}$-1$}%
}}}
\put(5356,-4786){\makebox(0,0)[lb]{\smash{\SetFigFont{12}{14.4}{\familydefault}{\mddefault}{\updefault}{\color[rgb]{0,0,0}$2$}%
}}}
\put(4006,-4786){\makebox(0,0)[lb]{\smash{\SetFigFont{12}{14.4}{\familydefault}{\mddefault}{\updefault}{\color[rgb]{0,0,0}$1$}%
}}}
\put(-134,-1456){\makebox(0,0)[lb]{\smash{\SetFigFont{12}{14.4}{\familydefault}{\mddefault}{\updefault}{\color[rgb]{0,0,0}$4$}%
}}}
\put(-134,-2266){\makebox(0,0)[lb]{\smash{\SetFigFont{12}{14.4}{\familydefault}{\mddefault}{\updefault}{\color[rgb]{0,0,0}$0$}%
}}}
\put(-179,-4786){\makebox(0,0)[lb]{\smash{\SetFigFont{12}{14.4}{\familydefault}{\mddefault}{\updefault}{\color[rgb]{0,0,0}$-2$}%
}}}
\put(-269,-3121){\makebox(0,0)[lb]{\smash{\SetFigFont{12}{14.4}{\familydefault}{\mddefault}{\updefault}{\color[rgb]{0,0,0}$-4$}%
}}}
\put(-269,-3976){\makebox(0,0)[lb]{\smash{\SetFigFont{12}{14.4}{\familydefault}{\mddefault}{\updefault}{\color[rgb]{0,0,0}$-8$}%
}}}
\put(-134,-961){\makebox(0,0)[lb]{\smash{\SetFigFont{12}{14.4}{\familydefault}{\mddefault}{\updefault}{\color[rgb]{0,0,0}$q$}%
}}}
\end{picture}
\end{center}
\end{figure}

\newpage

\section*{Figure 3}

\vspace*{3cm}

\begin{figure}[h]
\begin{center}
\begin{picture}(0,0)%
\includegraphics{malomed_fig3.pstex}%
\end{picture}%
\setlength{\unitlength}{4144sp}%
\begingroup\makeatletter\ifx\SetFigFont\undefined%
\gdef\SetFigFont#1#2#3#4#5{%
  \reset@font\fontsize{#1}{#2pt}%
  \fontfamily{#3}\fontseries{#4}\fontshape{#5}%
  \selectfont}%
\fi\endgroup%
\begin{picture}(5627,6646)(-181,-10487)
\put(361,-4381){\makebox(0,0)[lb]{\smash{\SetFigFont{8}{9.6}{\familydefault}{\mddefault}{\updefault}{\color[rgb]{0,0,0}$W$}%
}}}
\put(3241,-4336){\makebox(0,0)[lb]{\smash{\SetFigFont{8}{9.6}{\familydefault}{\mddefault}{\updefault}{\color[rgb]{0,0,0}$W$}%
}}}
\put(3061,-5596){\makebox(0,0)[lb]{\smash{\SetFigFont{8}{9.6}{\familydefault}{\mddefault}{\updefault}{\color[rgb]{0,0,0}$V_1$}%
}}}
\put(2161,-5731){\makebox(0,0)[lb]{\smash{\SetFigFont{8}{9.6}{\familydefault}{\mddefault}{\updefault}{\color[rgb]{0,0,0}$V_2$}%
}}}
\put( 91,-5551){\makebox(0,0)[lb]{\smash{\SetFigFont{8}{9.6}{\familydefault}{\mddefault}{\updefault}{\color[rgb]{0,0,0}$V_1$}%
}}}
\put(-93,-7549){\makebox(0,0)[lb]{\smash{\SetFigFont{8}{9.6}{\familydefault}{\mddefault}{\updefault}{\color[rgb]{0,0,0}$0$}%
}}}
\put(-93,-9924){\makebox(0,0)[lb]{\smash{\SetFigFont{8}{9.6}{\familydefault}{\mddefault}{\updefault}{\color[rgb]{0,0,0}$0$}%
}}}
\put(2849,-5209){\makebox(0,0)[lb]{\smash{\SetFigFont{8}{9.6}{\familydefault}{\mddefault}{\updefault}{\color[rgb]{0,0,0}$0$}%
}}}
\put(2845,-7617){\makebox(0,0)[lb]{\smash{\SetFigFont{8}{9.6}{\familydefault}{\mddefault}{\updefault}{\color[rgb]{0,0,0}$0$}%
}}}
\put(2845,-9915){\makebox(0,0)[lb]{\smash{\SetFigFont{8}{9.6}{\familydefault}{\mddefault}{\updefault}{\color[rgb]{0,0,0}$0$}%
}}}
\put(2849,-3925){\makebox(0,0)[lb]{\smash{\SetFigFont{8}{9.6}{\familydefault}{\mddefault}{\updefault}{\color[rgb]{0,0,0}$8$}%
}}}
\put(-79,-5218){\makebox(0,0)[lb]{\smash{\SetFigFont{8}{9.6}{\familydefault}{\mddefault}{\updefault}{\color[rgb]{0,0,0}$0$}%
}}}
\put(2749,-5843){\makebox(0,0)[lb]{\smash{\SetFigFont{8}{9.6}{\familydefault}{\mddefault}{\updefault}{\color[rgb]{0,0,0}$-4$}%
}}}
\put(-181,-5843){\makebox(0,0)[lb]{\smash{\SetFigFont{8}{9.6}{\familydefault}{\mddefault}{\updefault}{\color[rgb]{0,0,0}$-4$}%
}}}
\put(2849,-4565){\makebox(0,0)[lb]{\smash{\SetFigFont{8}{9.6}{\familydefault}{\mddefault}{\updefault}{\color[rgb]{0,0,0}$4$}%
}}}
\put(-83,-4565){\makebox(0,0)[lb]{\smash{\SetFigFont{8}{9.6}{\familydefault}{\mddefault}{\updefault}{\color[rgb]{0,0,0}$4$}%
}}}
\put(-80,-3925){\makebox(0,0)[lb]{\smash{\SetFigFont{8}{9.6}{\familydefault}{\mddefault}{\updefault}{\color[rgb]{0,0,0}$8$}%
}}}
\put(-181,-8082){\makebox(0,0)[lb]{\smash{\SetFigFont{8}{9.6}{\familydefault}{\mddefault}{\updefault}{\color[rgb]{0,0,0}$-4$}%
}}}
\put(2750,-8082){\makebox(0,0)[lb]{\smash{\SetFigFont{8}{9.6}{\familydefault}{\mddefault}{\updefault}{\color[rgb]{0,0,0}$-4$}%
}}}
\put(2750,-10339){\makebox(0,0)[lb]{\smash{\SetFigFont{8}{9.6}{\familydefault}{\mddefault}{\updefault}{\color[rgb]{0,0,0}$-4$}%
}}}
\put(-175,-10339){\makebox(0,0)[lb]{\smash{\SetFigFont{8}{9.6}{\familydefault}{\mddefault}{\updefault}{\color[rgb]{0,0,0}$-4$}%
}}}
\put(-94,-7001){\makebox(0,0)[lb]{\smash{\SetFigFont{8}{9.6}{\familydefault}{\mddefault}{\updefault}{\color[rgb]{0,0,0}$4$}%
}}}
\put(-94,-9499){\makebox(0,0)[lb]{\smash{\SetFigFont{8}{9.6}{\familydefault}{\mddefault}{\updefault}{\color[rgb]{0,0,0}$4$}%
}}}
\put(2834,-9496){\makebox(0,0)[lb]{\smash{\SetFigFont{8}{9.6}{\familydefault}{\mddefault}{\updefault}{\color[rgb]{0,0,0}$4$}%
}}}
\put(2842,-7137){\makebox(0,0)[lb]{\smash{\SetFigFont{8}{9.6}{\familydefault}{\mddefault}{\updefault}{\color[rgb]{0,0,0}$4$}%
}}}
\put(-91,-6434){\makebox(0,0)[lb]{\smash{\SetFigFont{8}{9.6}{\familydefault}{\mddefault}{\updefault}{\color[rgb]{0,0,0}$8$}%
}}}
\put(2847,-6656){\makebox(0,0)[lb]{\smash{\SetFigFont{8}{9.6}{\familydefault}{\mddefault}{\updefault}{\color[rgb]{0,0,0}$8$}%
}}}
\put(-81,-9063){\makebox(0,0)[lb]{\smash{\SetFigFont{8}{9.6}{\familydefault}{\mddefault}{\updefault}{\color[rgb]{0,0,0}$8$}%
}}}
\put(2844,-9063){\makebox(0,0)[lb]{\smash{\SetFigFont{8}{9.6}{\familydefault}{\mddefault}{\updefault}{\color[rgb]{0,0,0}$8$}%
}}}
\put(2773,-6171){\makebox(0,0)[lb]{\smash{\SetFigFont{8}{9.6}{\familydefault}{\mddefault}{\updefault}{\color[rgb]{0,0,0}$12$}%
}}}
\put(2773,-8639){\makebox(0,0)[lb]{\smash{\SetFigFont{8}{9.6}{\familydefault}{\mddefault}{\updefault}{\color[rgb]{0,0,0}$12$}%
}}}
\put(-152,-8639){\makebox(0,0)[lb]{\smash{\SetFigFont{8}{9.6}{\familydefault}{\mddefault}{\updefault}{\color[rgb]{0,0,0}$12$}%
}}}
\put(1576,-4111){\makebox(0,0)[lb]{\smash{\SetFigFont{8}{9.6}{\familydefault}{\mddefault}{\updefault}{\color[rgb]{0,0,0}$q=-2.8384$}%
}}}
\put(4501,-4111){\makebox(0,0)[lb]{\smash{\SetFigFont{8}{9.6}{\familydefault}{\mddefault}{\updefault}{\color[rgb]{0,0,0}$q=-2.2760$}%
}}}
\put(4501,-6361){\makebox(0,0)[lb]{\smash{\SetFigFont{8}{9.6}{\familydefault}{\mddefault}{\updefault}{\color[rgb]{0,0,0}$q=-1.3504$}%
}}}
\put(1576,-6361){\makebox(0,0)[lb]{\smash{\SetFigFont{8}{9.6}{\familydefault}{\mddefault}{\updefault}{\color[rgb]{0,0,0}$q=-1.4914$}%
}}}
\put(1576,-8611){\makebox(0,0)[lb]{\smash{\SetFigFont{8}{9.6}{\familydefault}{\mddefault}{\updefault}{\color[rgb]{0,0,0}$q=-1.0679$}%
}}}
\put(4501,-8611){\makebox(0,0)[lb]{\smash{\SetFigFont{8}{9.6}{\familydefault}{\mddefault}{\updefault}{\color[rgb]{0,0,0}$q=-1.0021$}%
}}}
\put(2251,-5956){\makebox(0,0)[lb]{\smash{\SetFigFont{8}{9.6}{\familydefault}{\mddefault}{\updefault}{\color[rgb]{0,0,0}$6$}%
}}}
\put( 46,-8206){\makebox(0,0)[lb]{\smash{\SetFigFont{8}{9.6}{\familydefault}{\mddefault}{\updefault}{\color[rgb]{0,0,0}$-6$}%
}}}
\put( 46,-10456){\makebox(0,0)[lb]{\smash{\SetFigFont{8}{9.6}{\familydefault}{\mddefault}{\updefault}{\color[rgb]{0,0,0}$-6$}%
}}}
\put(2971,-10456){\makebox(0,0)[lb]{\smash{\SetFigFont{8}{9.6}{\familydefault}{\mddefault}{\updefault}{\color[rgb]{0,0,0}$-6$}%
}}}
\put(2251,-8206){\makebox(0,0)[lb]{\smash{\SetFigFont{8}{9.6}{\familydefault}{\mddefault}{\updefault}{\color[rgb]{0,0,0}$6$}%
}}}
\put(2251,-10456){\makebox(0,0)[lb]{\smash{\SetFigFont{8}{9.6}{\familydefault}{\mddefault}{\updefault}{\color[rgb]{0,0,0}$6$}%
}}}
\put(2971,-8206){\makebox(0,0)[lb]{\smash{\SetFigFont{8}{9.6}{\familydefault}{\mddefault}{\updefault}{\color[rgb]{0,0,0}$-6$}%
}}}
\put(2971,-5956){\makebox(0,0)[lb]{\smash{\SetFigFont{8}{9.6}{\familydefault}{\mddefault}{\updefault}{\color[rgb]{0,0,0}$-6$}%
}}}
\put(5176,-10456){\makebox(0,0)[lb]{\smash{\SetFigFont{8}{9.6}{\familydefault}{\mddefault}{\updefault}{\color[rgb]{0,0,0}$6$}%
}}}
\put(5176,-8206){\makebox(0,0)[lb]{\smash{\SetFigFont{8}{9.6}{\familydefault}{\mddefault}{\updefault}{\color[rgb]{0,0,0}$6$}%
}}}
\put(5176,-5956){\makebox(0,0)[lb]{\smash{\SetFigFont{8}{9.6}{\familydefault}{\mddefault}{\updefault}{\color[rgb]{0,0,0}$6$}%
}}}
\put( 46,-5956){\makebox(0,0)[lb]{\smash{\SetFigFont{8}{9.6}{\familydefault}{\mddefault}{\updefault}{\color[rgb]{0,0,0}$-6$}%
}}}
\put(766,-5956){\makebox(0,0)[lb]{\smash{\SetFigFont{8}{9.6}{\familydefault}{\mddefault}{\updefault}{\color[rgb]{0,0,0}$-2$}%
}}}
\put(1531,-5956){\makebox(0,0)[lb]{\smash{\SetFigFont{8}{9.6}{\familydefault}{\mddefault}{\updefault}{\color[rgb]{0,0,0}$\, 2$}%
}}}
\put(1531,-8206){\makebox(0,0)[lb]{\smash{\SetFigFont{8}{9.6}{\familydefault}{\mddefault}{\updefault}{\color[rgb]{0,0,0}$\, 2$}%
}}}
\put(1531,-10456){\makebox(0,0)[lb]{\smash{\SetFigFont{8}{9.6}{\familydefault}{\mddefault}{\updefault}{\color[rgb]{0,0,0}$\, 2$}%
}}}
\put(766,-8206){\makebox(0,0)[lb]{\smash{\SetFigFont{8}{9.6}{\familydefault}{\mddefault}{\updefault}{\color[rgb]{0,0,0}$-2$}%
}}}
\put(766,-10456){\makebox(0,0)[lb]{\smash{\SetFigFont{8}{9.6}{\familydefault}{\mddefault}{\updefault}{\color[rgb]{0,0,0}$-2$}%
}}}
\put(3691,-5956){\makebox(0,0)[lb]{\smash{\SetFigFont{8}{9.6}{\familydefault}{\mddefault}{\updefault}{\color[rgb]{0,0,0}$-2$}%
}}}
\put(3691,-8206){\makebox(0,0)[lb]{\smash{\SetFigFont{8}{9.6}{\familydefault}{\mddefault}{\updefault}{\color[rgb]{0,0,0}$-2$}%
}}}
\put(3691,-10456){\makebox(0,0)[lb]{\smash{\SetFigFont{8}{9.6}{\familydefault}{\mddefault}{\updefault}{\color[rgb]{0,0,0}$-2$}%
}}}
\put(4456,-5956){\makebox(0,0)[lb]{\smash{\SetFigFont{8}{9.6}{\familydefault}{\mddefault}{\updefault}{\color[rgb]{0,0,0}$\, 2$}%
}}}
\put(4456,-8206){\makebox(0,0)[lb]{\smash{\SetFigFont{8}{9.6}{\familydefault}{\mddefault}{\updefault}{\color[rgb]{0,0,0}$\, 2$}%
}}}
\put(4456,-10456){\makebox(0,0)[lb]{\smash{\SetFigFont{8}{9.6}{\familydefault}{\mddefault}{\updefault}{\color[rgb]{0,0,0}$\, 2$}%
}}}
\put(5131,-4696){\makebox(0,0)[lb]{\smash{\SetFigFont{8}{9.6}{\familydefault}{\mddefault}{\updefault}{\color[rgb]{0,0,0}$V_2$}%
}}}
\put(2521,-5776){\makebox(0,0)[lb]{\smash{\SetFigFont{10}{12.0}{\familydefault}{\mddefault}{\updefault}{\color[rgb]{0,0,0}$x$}%
}}}
\put(5446,-5776){\makebox(0,0)[lb]{\smash{\SetFigFont{10}{12.0}{\familydefault}{\mddefault}{\updefault}{\color[rgb]{0,0,0}$x$}%
}}}
\put(2521,-8026){\makebox(0,0)[lb]{\smash{\SetFigFont{10}{12.0}{\familydefault}{\mddefault}{\updefault}{\color[rgb]{0,0,0}$x$}%
}}}
\put(5446,-8026){\makebox(0,0)[lb]{\smash{\SetFigFont{10}{12.0}{\familydefault}{\mddefault}{\updefault}{\color[rgb]{0,0,0}$x$}%
}}}
\put(5446,-10276){\makebox(0,0)[lb]{\smash{\SetFigFont{10}{12.0}{\familydefault}{\mddefault}{\updefault}{\color[rgb]{0,0,0}$x$}%
}}}
\put(2521,-10276){\makebox(0,0)[lb]{\smash{\SetFigFont{10}{12.0}{\familydefault}{\mddefault}{\updefault}{\color[rgb]{0,0,0}$x$}%
}}}
\end{picture}

\end{center}
\end{figure}

\newpage

\section*{Figure 4}

\vspace*{2cm}

\begin{figure}[h]
\begin{center}
\begin{picture}(0,0)%
\includegraphics{malomed_fig4.pstex}%
\end{picture}%
\setlength{\unitlength}{4144sp}%
\begingroup\makeatletter\ifx\SetFigFont\undefined%
\gdef\SetFigFont#1#2#3#4#5{%
  \reset@font\fontsize{#1}{#2pt}%
  \fontfamily{#3}\fontseries{#4}\fontshape{#5}%
  \selectfont}%
\fi\endgroup%
\begin{picture}(6333,7204)(2206,-10406)
\put(6391,-4786){\makebox(0,0)[lb]{\smash{\SetFigFont{10}{12.0}{\familydefault}{\mddefault}{\updefault}{\color[rgb]{0,0,0}b) $n=2$}%
}}}
\put(6391,-10006){\makebox(0,0)[lb]{\smash{\SetFigFont{10}{12.0}{\familydefault}{\mddefault}{\updefault}{\color[rgb]{0,0,0}d) $n=20$}%
}}}
\put(4521,-6786){\makebox(0,0)[lb]{\smash{\SetFigFont{10}{12.0}{\familydefault}{\mddefault}{\updefault}{\color[rgb]{0,0,0}b)}%
}}}
\put(4516,-5831){\makebox(0,0)[lb]{\smash{\SetFigFont{10}{12.0}{\familydefault}{\mddefault}{\updefault}{\color[rgb]{0,0,0}a)}%
}}}
\put(3226,-7651){\makebox(0,0)[lb]{\smash{\SetFigFont{10}{12.0}{\familydefault}{\mddefault}{\updefault}{\color[rgb]{0,0,0}f)}%
}}}
\put(2906,-7121){\makebox(0,0)[lb]{\smash{\SetFigFont{10}{12.0}{\familydefault}{\mddefault}{\updefault}{\color[rgb]{0,0,0}c)}%
}}}
\put(2446,-7036){\makebox(0,0)[lb]{\smash{\SetFigFont{10}{12.0}{\familydefault}{\mddefault}{\updefault}{\color[rgb]{0,0,0}d)}%
}}}
\put(5446,-7846){\makebox(0,0)[lb]{\smash{\SetFigFont{12}{14.4}{\familydefault}{\mddefault}{\updefault}{\color[rgb]{0,0,0}$n^{-6/5}$}%
}}}
\put(2251,-5641){\makebox(0,0)[lb]{\smash{\SetFigFont{12}{14.4}{\familydefault}{\mddefault}{\updefault}{\color[rgb]{0,0,0}$\varepsilon$}%
}}}
\put(3001,-8656){\makebox(0,0)[lb]{\smash{\SetFigFont{10}{12.0}{\familydefault}{\mddefault}{\updefault}{\color[rgb]{0,0,0}f)}%
}}}
\put(2476,-5686){\makebox(0,0)[lb]{\smash{\SetFigFont{12}{14.4}{\familydefault}{\mddefault}{\updefault}{\color[rgb]{0,0,0}e)}%
}}}
\put(2971,-4831){\makebox(0,0)[lb]{\smash{\SetFigFont{10}{12.0}{\familydefault}{\mddefault}{\updefault}{\color[rgb]{0,0,0}a) $n=1$}%
}}}
\put(3241,-3481){\makebox(0,0)[lb]{\smash{\SetFigFont{10}{12.0}{\familydefault}{\mddefault}{\updefault}{\color[rgb]{0,0,0}$v$}%
}}}
\put(4411,-4786){\makebox(0,0)[lb]{\smash{\SetFigFont{10}{12.0}{\familydefault}{\mddefault}{\updefault}{\color[rgb]{0,0,0}$w$}%
}}}
\put(6391,-7441){\makebox(0,0)[lb]{\smash{\SetFigFont{10}{12.0}{\familydefault}{\mddefault}{\updefault}{\color[rgb]{0,0,0}c) $n=10$}%
}}}
\put(3891,-5141){\makebox(0,0)[lb]{\smash{\SetFigFont{10}{12.0}{\familydefault}{\mddefault}{\updefault}{\color[rgb]{0,0,0}$x$}%
}}}
\put(7311,-5141){\makebox(0,0)[lb]{\smash{\SetFigFont{10}{12.0}{\familydefault}{\mddefault}{\updefault}{\color[rgb]{0,0,0}$x$}%
}}}
\put(7321,-7756){\makebox(0,0)[lb]{\smash{\SetFigFont{10}{12.0}{\familydefault}{\mddefault}{\updefault}{\color[rgb]{0,0,0}$x$}%
}}}
\put(7331,-10366){\makebox(0,0)[lb]{\smash{\SetFigFont{10}{12.0}{\familydefault}{\mddefault}{\updefault}{\color[rgb]{0,0,0}$x$}%
}}}
\end{picture}
\end{center}
\end{figure}

\end{document}